\documentstyle[twoside,fleqn,espcrc2,epsf]{article}

\newcommand{\AmS}{{\protect\the\textfont2
  A\kern-.1667em\lower.5ex\hbox{M}\kern-.125emS}}

\def\eq#1{Eq.~(\ref{#1})}

\def\tab#1{Table \ref{#1}}

\def\fig#1{Fig.~\ref{#1}}

\def\l{\left}
\def\r{\right}
\def\la{\langle}
\def\ra{\rangle}

\def\nn{\nonumber}

\def\beq{\begin{equation}}
\def\eeq{\end{equation}}
\def\bea{\begin{eqnarray}}
\def\eea{\end{eqnarray}}

\def\gev{\mbox{GeV}}

\def\msbar{{\overline{\mathrm{MS}}}}

\title{$\Delta S=2$ and $\Delta I=3/2$ Matrix Elements in Quenched QCD
\thanks{CPT-98/PE.3690; CERN-TH/98-307. 
Support from EPSRC and PPARC under grants GR/K41663 and GR/L29927, and
from the EEC through TMR network EEC-CT98-00169
is acknowledged.}}

\author{Laurent Lellouch
        \address{Theory Division, CERN, CH-1211 Geneva 23,
        Switzerland}
        \thanks{Presenter; 
          on leave from Centre de Physique Th\'eorique, Case
        907, CNRS Luminy, F-13288 Marseille Cedex 9
.}%
        and 
        C.-J. David Lin\address{Department of Physics $\&$ Astronomy, 
        The University of Edinburgh, Edinburgh EH9 3JZ, Scotland}
        (UKQCD Collaboration)}
       
\begin{document}

\begin{abstract}
  We present preliminary results for $B_K$, $B_7^{3/2}$ and
  $B_8^{3/2}$ from two high-statistics lattice computations. These
  calculations are performed at $\beta=6.0$ and 6.2 in the quenched
  approximation, using mean-field-improved Sheikholeslami-Wohlert
  fermionic actions.
\end{abstract}

\maketitle

\section{Introduction}

We report on two, high-statistics, quenched lattice calculations of
$B_K$, $B_7^{3/2}$ and $B_8^{3/2}$. $B_K$ parametrizes the $K^0-\bar
K^0$-oscillation contribution to CP-violation in $K$ decays
($\epsilon$), leading to a hyperbolic constraint on the summit of the
unitarity triangle. $B_{7,8}^{3/2}$ are needed to compute the
electro-penguin contribution to direct CP-violation ($\epsilon'$),
dominant in the $\Delta I=3/2$ channel.  $B_K$ and $B_{7,8}^{3/2}$
measure deviation from the vacuum saturation values of the following
four-quark, matrix elements:
\[
\la\bar K^0|O_{\Delta S=2}|K^0\ra=
\frac{8}{3}|\la 0|\bar s\gamma_\mu\gamma^5 d|K^0\ra|^2\, B_K
\ ,
\]
with 
$O_{\Delta S=2}=(\bar s\gamma_\mu^L d)(\bar s\gamma^\mu_L d)$, where
$\gamma^\mu_{R,L}=1\pm\gamma_5$, 
and in the chiral limit, 
\[
\la\pi^+|O_7^{3/2}|K^+\ra \to
\frac{2}{3}\la\pi^+|\bar u\gamma^5 d|0\ra\la 0|\bar s\gamma^5 u|K^+\ra
\,B_7^{3/2}
\]
\[
\la\pi^+|O_8^{3/2}|K^+\ra  \to
2\la\pi^+|\bar u\gamma^5 d|0\ra\la 0|\bar s\gamma^5 u|K^+\ra
\,B_8^{3/2},
\]
where $O_7^{3/2}$ can be written as $(\bar
s\gamma_\mu^L d) (\bar u\gamma^\mu_R u)+ (\bar s\gamma_\mu^L u)(\bar
u\gamma^\mu_R d)$ if one forbids penguin contractions and where
$O_8^{3/2}$ is the corresponding color-mixed operator.

\section{Simulation Details}

We describe quarks with the mean-field-improved,
Sheikholeslami-Wohlert (SW) action
\[
  S_F^{SW} = S_F^W - ig\,c_{SW}\frac{\kappa}{2}\sum_{x,\mu,\nu}\,
             \bar q(x)\,P_{\mu\nu}\sigma_{\mu\nu}\,q(x)
\ ,
\label{eq:sw}
\]
with $c_{SW}=1/u_0^3$ and $u_0\equiv\la \frac{1}{3}{\rm Tr} U_{pl}
\ra^{\frac{1}{4}}$ and where $S_F^W$ is the Wilson fermion action,
$g$, the bare gauge coupling, $P_{\mu\nu}$, a lattice definition of
the field-strength tensor and $\kappa$, the appropriate quark hopping
parameter.  The parameters of the simulation are summarized in
\tab{tab:simparam}. We further perform full tadpole-improved, 
KLM rotation of quark fields. 
Though these normalization 
factors cancel in the calculation of $B$-parameters,
they specify our renomalization constants.

\begin{table*}[hbt]
\setlength{\tabcolsep}{1.5pc}
\newlength{\digitwidth} \settowidth{\digitwidth}{\rm 0}
\catcode`?=\active \def?{\kern\digitwidth}
\caption{
{\small\it 
Simulation parameters. The masses below each $\kappa$ are the corresponding
pseudoscalar-meson masses.}}
\label{tab:simparam}
\begin{tabular*}{\textwidth}{@{}l@{\extracolsep{\fill}}ccccccc}
\hline\hline
$\beta$ & size & $\#$ cfs. & $c_{SW}$ &
\multicolumn{3}{c}{$\kappa$}\\
\hline
6.2 & $24^3\times 48$ & 188 & 1.442 & 
0.13640 & 0.13710 & 0.13745\\
\multicolumn{4}{c}{$a^{-1}(m_\rho)=2.56^{+8}_{-8}\ \gev$}
& 780 MeV & 570 MeV & 438 MeV\\
6.0 & $16^3\times 48$ & 498 & 1.479 & 
0.13700 & 0.13810 & 0.13856\\
\multicolumn{4}{c}{$a^{-1}(m_\rho)=1.96^{+6}_{-5}\ \gev$} 
& 811 MeV & 575 MeV & 445 MeV\\
\hline\hline
\vspace{-24pt}
\end{tabular*}
\end{table*}

\section{Operator Matching}

While the matching of quark bilinears is simple, the use of Wilson
fermions induces mixing amongst four-quark operators of different
chirality. To describe this mixing, we use the following complete
basis of parity-conserving operators:
\[
\begin{array}{ccc}
O_{1,2}^{lat} 
&=& \gamma_\mu\times\gamma_\mu\pm\gamma_\mu\gamma_5\times\gamma_\mu
\gamma_5\ ,\\
O_{3,4}^{lat} &=& I\times I \pm \gamma_5\times\gamma_5\ ,\\
O_5^{lat} &=& \sigma_{\mu\nu}\times\sigma_{\mu\nu}\ ,
\end{array}
\]
with $\Gamma\times\Gamma\equiv (\bar q'\Gamma q)(\bar q'\Gamma
q)^{lat}(a)$. Then, the matching can be written, with 
identifications that will be clarified by what follows,
\[
O_{\Delta F=2}(\mu) \to
\hat Z_{11}\ \hat O_1^{lat}(a)
\]
and
\[
\l(
\begin{array}{c}
O_7^{3/2}(\mu) \\
-O_8^{3/2}(\mu)/2
\end{array}
\r)
\to
\l(
\begin{array}{cc}
\hat Z_{22} & \hat Z_{24}\\
\hat Z_{42} & \hat Z_{44}
\end{array}
\r)
\l(
\begin{array}{c}
\hat O_2^{lat}(a) \\
\hat O_4^{lat}(a)
\end{array}
\r)
\]
in terms of the chirally subtracted operators $\hat O_1^{lat}(a)\equiv
O_1^{lat}(a)+ \sum_{i=2}^5 Z_{1i}\, O_i^{lat}(a)$ and
$\hat O_{i}^{lat}(a)\equiv O_{i}^{lat}(a)+\sum_{j=1,3,5} 
Z_{ij}\,O_i^{lat}(a)$, i=2,4.

We perform matching to the $\msbar$-NDR scheme at one
loop. Since the clover term is ${\mathcal{O}}(g)$, we can use the
tree-level-clover-action results of \cite{RFr92} with modifications
appropriate for tadpole-improvement and KLM normalization. For the
coupling, we choose $\alpha_s^\msbar(\mu)$ defined from the
plaquette~\cite{GLe93}, identifying the scale $\mu$ with that of the
matching. To estimate the systematic error associated with our
procedure, we vary $\mu$ in the range $1/a\to\pi/a$.

\section{Analysis and Results}

To obtain the desired $B$-parameters, we consider ratios of 3-point 
to two 2-point
functions. In the limit that the three points are well separated in time,
these ratios reduce to:
\bea
R_{\Delta F=2} &\rightarrow& \frac{1}{Z_{\gamma_\mu\gamma_5}^2}
       \frac{\la\bar P(\vec{q})|
       O_1^{(NDR)} |P(\vec{p})\ra}
       {|\la 0|P^{lat}|P\ra|^2}\nn\\
R_7^{3/2} &\rightarrow& -\frac{3}{4}\frac{\la\bar P(\vec{q})|
       O_2^{(NDR)} |P(\vec{p})\ra}
       {Z_{\gamma_5}^2\,|\la 0|P^{lat}|P\ra|^2}
\nn\label{eq:ratdefs}
\\
R_8^{3/2} &\rightarrow& \frac{1}{2}
       \frac{\la\bar P(\vec{q})|O_4^{(NDR)} |P(\vec{p})\ra}
       {Z_{\gamma_5}^2\,|\la 0|P^{lat}|P\ra|^2}\nn,
\eea
where $P$ is a $q\bar q'$ pseudoscalar meson. We calculate the
ratios for the momenta
$\vec{p}\to\vec{q} = 0\to 0$, $0\to 1$, $1\to 1_\perp$ and $1\to
1_\parallel$ and for all hopping parameter pairs taken from
\tab{tab:simparam}.

To study their chiral behavior, we follow \cite{MCr96} and define the mass
and recoil variables
\bea
X &= & \frac{8}{3}\frac{(f_P^{lat})^2 M_P^2}{|\la 0|
       P^{lat}|P\ra|^2}\nn\\
Y &=& \frac{p\cdot q}{M_P^2}X. \nn
\label{eq:xydef}
\eea
We then fit the ratios to
\beq
R(X,Y) = a_{00}+a_{10}\,X+a_{01}\,Y+\cdots
\label{eq:rxy}
\ ,
\eeq
where we neglect both chiral logarithms, which are difficult to 
resolve numerically, and $SU(3)_f$ breaking terms, which appear to be small
for the quark masses we consider.

\begin{figure*}[htb]
\centerline{\epsfxsize=12cm\epsffile{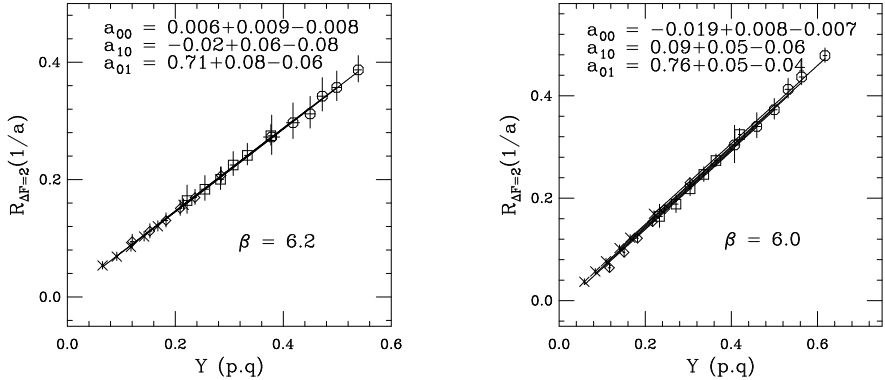}}
\vspace{-28pt}
\caption{\small\it
Chiral behavior of $R_{\Delta F=2}$ for $\mu=1/a$. The different symbols 
correspond to different $\vec{p}\to\vec{q}$ while the different points
within a given set correspond to different pseudoscalar meson masses.
\vspace{-24pt}
}
\label{fig:chiralbk}
\end{figure*}

We find that $R_{\Delta F=2}$ is well described by a linear form in $X$ and 
$Y$ as shown in \fig{fig:chiralbk} for $\mu=1/a$.
At $\beta=6.2$ we further find that $a_{00}$ and $a_{10}$ are very
small and consistent with zero, as chiral symmetry requires, and
remain so as $\mu$ is increased up to $\pi/a$. $a_{01}$ should thus
give a reliable estimate of $B_K$. At $\beta=6.0$, $a_{00}$ and
$a_{10}$ are less than 3 and $2\sigma$ away from zero and smaller than
the values obtained in \cite{LCo98} with a tree-level SW action and
boosted, one-loop matching.  Taken in conjunction with these and other
Wilson results obtained from less improved actions, our preliminary
findings suggest that discretization errors represent an important
part of the traditionally observed residual chiral violations (please
see \cite{RGu98} and \cite{LCo98,MCr96} for further discussion).

In the chiral limit, $B_{7,8}^{3/2}$ correspond to the leading term,
$a_{00}$, in the expansion of \eq{eq:rxy} and the study of the
$Y$-dependence of $R_{7,8}^{3/2}$ is less important. Therefore, we
consider only the $\vec{p}= \vec{q}=\vec{0}$ ratios so as not to
introduce potential momentum-dependent discretization errors. We
find that the description of the chiral behavior of $R_{7,8}^{3/2}$
requires a quadratic term in $X$.

To compare results for $B_K$ and $B_{7,8}^{3/2}$ obtained at different
$\mu$ and/or $\beta$, we must run them to a common reference scale
which we take to be $2\,\gev$. Running is performed at two loops with
$n_f{=}0$.  (See \tab{tab:bk782gev}.)
\begin{table}
\begin{center}
\caption{\small\it $B$-parameters at 2 GeV in the $\msbar$-NDR scheme
as a function of $\beta$ and the matching scale, $\mu$.}
\label{tab:bk782gev}
\begin{tabular}{ccccc}
\hline\hline
$\beta$ & $\mu$ & $B_K$ & $B^{3/2}_7$ & $B^{3/2}_8$\\
\hline
6.2 & $\pi/a$ & $0.71^{+8}_{-6}$ & $0.60^{+5}_{-4}$ & $0.81^{+8}_{-8}$\\
 & $2/a$ & $0.71^{+8}_{-6}$ & $0.58^{+5}_{-4}$ & $0.80^{+8}_{-8}$\\
 & $1/a$ & $0.72^{+8}_{-6}$ & $0.50^{+4}_{-4}$ & $0.76^{+8}_{-8}$\\
\hline
6.0 & $\pi/a$ & $0.75^{+4}_{-4}$ & $0.55^{+3}_{-3}$ & $0.74^{+3}_{-3}$\\
 & $2/a$ & $0.75^{+4}_{-4}$ & $0.53^{+3}_{-3}$ & $0.73^{+3}_{-3}$\\
 & $1/a$ & $0.76^{+5}_{-4}$ & $0.43^{+3}_{-2}$ & $0.68^{+3}_{-3}$\\
\hline\hline
\vspace{-42pt}
\end{tabular}
\end{center}
\end{table}

We find that $B_K$ is almost independent of $\mu$ in the range
explored. $a$-dependence is also found to be small though it
cannot be excluded given the statistical errors and the small deviations of
$a_{00}$ and $a_{10}$ from zero at 6.0. Two values of the lattice
spacing are not sufficient for a proper continuum-limit extrapolation
and we quote as our preliminary 
result for $B_K$ the $\mu=1/a$ number at $\beta=6.2$,
with the comment that residual discretization errors may be small.

The $\mu$-dependence of $B_8^{3/2}$ and especially $B_7^{3/2}$ is
significantly stronger than that of $B_K$. This is a result of the
rather large matching constants and anomalous dimensions. We favor the
results obtained at the larger values of $\mu$ because the matching
times running is much better behaved than at $\mu=1/a$.  Though
$a$-dependence is not, on the whole, significant statistically at
fixed $a\mu$ or $\mu$, discretization errors are difficult to quantify
because of the large $\mu$-dependence.  Once again we cannot
extrapolate to the continuum limit, so we quote as our preliminary
results for $B_{7,8}^{3/2}$ the $\mu=2/a$ numbers
at $\beta=6.2$ to which we add systematic errors to account for the
observed $\mu$-dependence.

Our results are summarized in \tab{tab:finalres}. Further discussion of
systematic uncertainties must
be postponed for lack of space.  For further comparisons
with other recent results, please see\cite{guido}.
\begin{table}
\begin{center}
\caption{\small\it Preliminary results for 
$B$-parameters at 2~GeV in the $\msbar$-NDR scheme
(see text).}
\label{tab:finalres}
\begin{tabular}{ccc}
\hline\hline
$B_K$ & $B^{3/2}_7$ & $B^{3/2}_8$\\
\hline
$0.72^{+8}_{-6}$ & $0.58^{+5+2}_{-4-8}$ & $0.80^{+8+1}_{-8-4}$\\
\hline\hline
\vspace{-42pt}
\end{tabular}
\end{center}
\end{table}

\end{document}